\documentclass[trackchanges]{aastex7}
\usepackage{stfloats}
\usepackage{caption}

\begin{document}

\title{Observational Properties of $\beta$ Cephei Stars: 88 new samples discovered Based on TESS and Gaia Data}

\author[orcid=0000-0002-5038-5952]{Xiang-dong Shi}
\affiliation{Yunnan Observatories, Chinese Academy of Sciences(CAS), P.O. Box 110, Kunming 650216, People’s Republic of China}
\affiliation{University of Chinese Academy of Sciences, No.1 Yanqihu East Rd, Huairou District, Beijing 101408, People’s Republic of China}
\email{sxd@ynao.ac.cn}  

\author{Sheng-bang Qian} 
\affiliation{Department of Astronomy, Key Laboratory of Astroparticle Physics of Yunnan Province, Yunnan University, Kunming 650091, People’s Republic of China}
\email{qiansb@ynu.edu.cn}

\author{Li-ying Zhu}
\affiliation{Yunnan Observatories, Chinese Academy of Sciences(CAS), P.O. Box 110, Kunming 650216, People’s Republic of China}
\affiliation{University of Chinese Academy of Sciences, No.1 Yanqihu East Rd, Huairou District, Beijing 101408, People’s Republic of China}
\email[show]{zhuly@ynao.ac.cn}

\author{Lin-jia Li}
\affiliation{Yunnan Observatories, Chinese Academy of Sciences(CAS), P.O. Box 110, Kunming 650216, People’s Republic of China}
\affiliation{University of Chinese Academy of Sciences, No.1 Yanqihu East Rd, Huairou District, Beijing 101408, People’s Republic of China}
\email{lipk@ynao.ac.cn}

\author{Er-gang Zhao}
\affiliation{Yunnan Observatories, Chinese Academy of Sciences(CAS), P.O. Box 110, Kunming 650216, People’s Republic of China}
\affiliation{University of Chinese Academy of Sciences, No.1 Yanqihu East Rd, Huairou District, Beijing 101408, People’s Republic of China}
\email{zergang@ynao.ac.cn}

\author{David Mkrtichian}
\affiliation{National Astronomical Research Institute of Thailand, 260 Moo 4, T. Donkaew, A. Maerim 50180, Chiangmai, Thailand}
\email{davidmkrt@gmail.com}

\author{Farkhodjon Khamrakulov}
\affiliation{Samarkand State University named after Sharof Rashidov, Engineering Physics Institute, Department of Nuclear Physics and Astronomy, University blv. 15, 140104 Samarkand, Uzbekistan}
\email{x-farxodjon@mail.ru}

\author{Wen-xu Lin}
\affiliation{Yunnan Observatories, Chinese Academy of Sciences(CAS), P.O. Box 110, Kunming 650216, People’s Republic of China}
\affiliation{University of Chinese Academy of Sciences, No.1 Yanqihu East Rd, Huairou District, Beijing 101408, People’s Republic of China}
\email{linwenxu@ynao.ac.cn}

\begin{abstract}

We present a systematic investigation of $\beta$ Cephei (BCEP) stars by integrating photometric data from the Transiting Exoplanet Survey Satellite (TESS) with astrometric parameters from Gaia Data Release 3. Utilizing TESS's short-cadence (SC) and full-frame image (FFI) photometry, along with Gaia parallaxes and temperatures derived from the Extended Stellar Parametrizer for Hot Stars (ESP-HS) pipeline, we identify 88 new BCEP stars and candidates--85 from SC data and 3 from SPOC-processed FFI observations. These targets exhibit visual magnitudes ranging from 8.0 to 12.0 mag, parallaxes between 0.11 and 1.74 mas, effective temperatures of 18,000 to 30,000 K, and luminosities from 1,500--38,000 $L_\odot$, consistent with previously cataloged BCEP populations, thereby demonstrating the robustness of our classification criteria. Key findings include: (1) a significant detection disparity between SC and FFI datasets, with 30\% of SC targets exceeding 18,000 K compared to only 0.7\% in FFI, reflecting observational biases toward high-luminosity, hotter stars in SC data; (2) four samples near the red edge of the theoretical instability strip, exhibiting sparse pulsation modes that are important samples for testing pulsation models under low-mass, low-temperature conditions; and (3) spatial clustering within the Galactic disk ($|b| < 20^\circ$), with two high-latitude outliers likely representing runaway stars ejected from disk environments. Our analysis underscores the critical role of space-based photometry in detecting low-amplitude pulsators and the transformative potential of multi-survey integration in the era of time-domain astronomy. These results provide new samples to constrain stellar pulsation theories of massive stars and to study Galactic dynamics.

\end{abstract}



\section{Introduction}
Massive stars, which serve as progenitors of compact objects such as neutron stars and black holes, are the source of extreme astrophysical events, including supernova explosions, mergers of compact binaries, and gravitational wave radiation \citep{2008ApJ...676.1162S, 2010ApJ...725..940Y, 2020RAA....20..161H, 2020A&A...638A..39L}. Simultaneously, these stars play a crucial role in the chemical enrichment of galaxies through supernova explosions and stellar wind mass loss, which directly influence the distribution of metal abundance in the interstellar medium \citep{2001A&A...369..574V}. In the field of stellar evolution, massive pulsating variables--with their unique oscillatory properties--serve as essential probes for asteroseismology, helping to unravel stellar internal structures and evolutionary laws (e.g., \citet{2003Sci...300.1926A, 2006A&A...459..589M, 2006ApJ...642..470A, 2006MNRAS.365..327H, 2008MNRAS.385.2061D, 2012MNRAS.427..483B}). Currently, only two classes of massive pulsators are recognized on upper main sequence stars: Slowly Pulsating B-type (SPB) stars and $\beta$ Cephei (BCEP) stars. Among these, BCEP stars are considered a "key laboratory" for studying massive star evolution due to their distinctive physical properties.

Compared to SPB stars, BCEP stars occupy a higher mass range. Their lower mass limit is generally established at $7\,M_\odot$, while the upper limit remains uncertain, potentially extending beyond \(20\,M_\odot\). This corresponds to late O-type to early B-type spectral classifications \citep{2007CoAst.151...48M}. BCEP stars represent the most massive group of standing-wave pulsators discovered on the main sequence. Their pulsations are driven by the $\kappa$-mechanism in the ionization zone of iron-group elements \citep{1992A&A...256L...5M, 1993MNRAS.262..204D}, predominantly exhibiting low-order pressure modes (p-modes) and often including gravity modes (g-modes). The most prominent pulsation periods range from 2 to 7 hours \citep{2005ApJS..158..193S, 2010aste.book.....A}.

Since p-modes and g-modes traverse the radiative envelope and convective core of a star, respectively, their oscillation spectra encode multidimensional structural information from the core to the envelope (e.g., \citet{2006CoAst.147....6K}. This provides asteroseismology with unique, multi-layered diagnostic capabilities, offering invaluable advantages in constraining parameters such as the size of convective cores in massive stars, convective core overshooting, internal rotation profiles, and chemical stratification (e.g., \citet{2021NatAs...5..715P, 2023NatAs...7..913B}).

Despite their astrophysical significance, only 400 to 500 Galactic BCEP stars have been cataloged (e.g., \citet{2005ApJS..158..193S}, \citet{2008A&A...477..917P}, \citet{2019MNRAS.489.1304B}, \citet{2020AJ....160...32L}, \citet{2020MNRAS.493.5871B}). The advent of large-scale astronomical surveys, including the Transiting Exoplanet Survey Satellite (TESS; \citet{2015JATIS...1a4003R}) and the Gaia mission \citep{2016A&A...595A...1G, 2018A&A...616A...1G, 2021A&A...649A...1G}, has revolutionized the systematic investigation of variable stars through extensive, high-precision datasets (e.g., \citet{2023ApJS..265...33S, 2023ApJS..268...16S, 2024ApJS..271...28S}, hereafter referred to as Papers I, II, and III). For BCEP stars--characterized by low-amplitude pulsations--the continuous and high-precision time-series photometry from space-based observatories provides unparalleled capabilities for their detection and characterization (e.g., Papers I and III, \citet{2020MNRAS.493.5871B, 2024ApJS..272...25E}). By leveraging data from TESS and Gaia, this study presents new discoveries of BCEP stars and systematically analyzes their observational properties.

\begin{figure*}[hp!]
	\centering
	\plotone{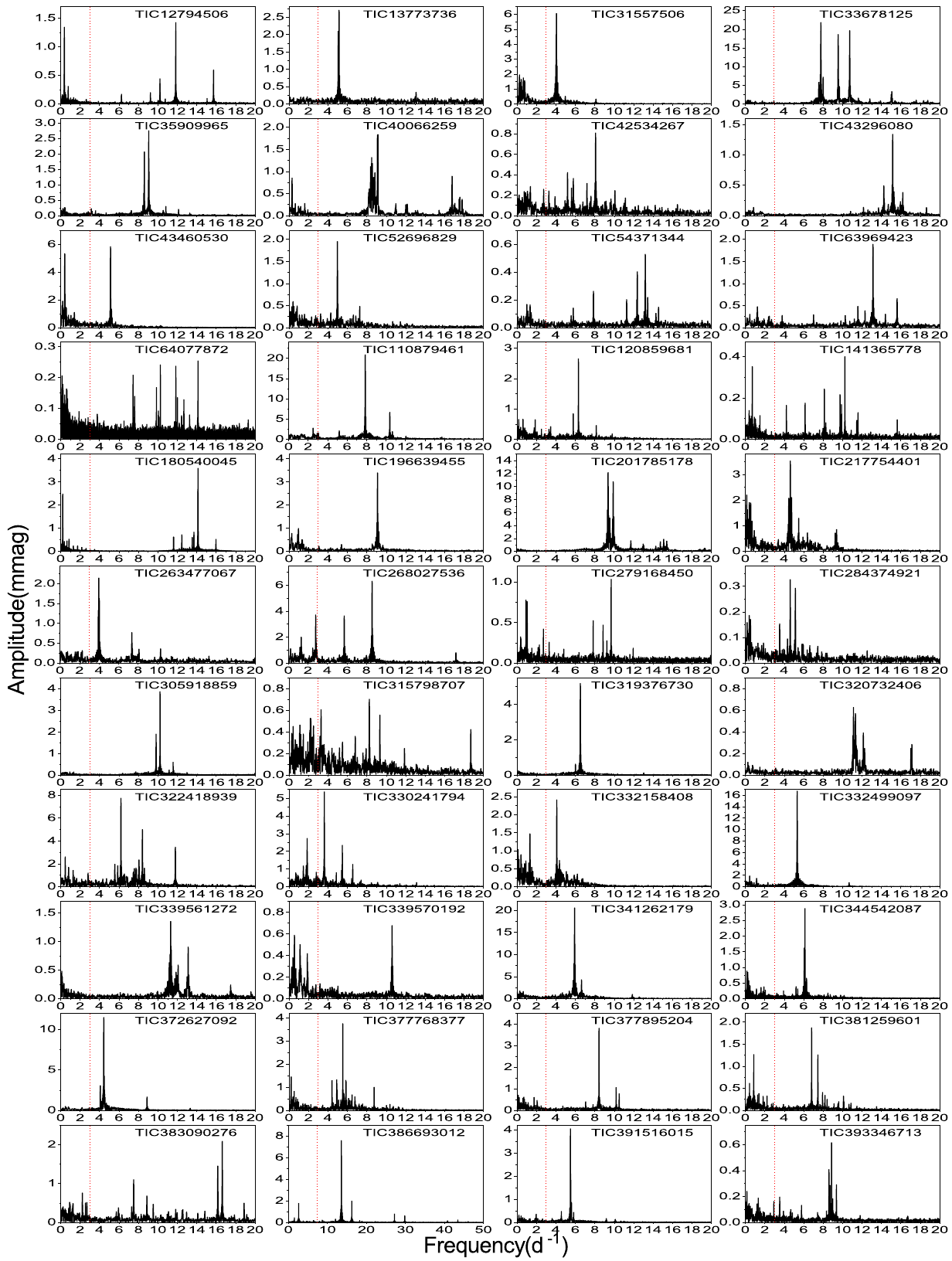}
	\caption{Fourier spectra of new BCEP stars and candidates. The red dotted lines are the rough boundary (3 $d^{-1}$) for low-frequency and high-frequency.}
	\label{fig:FS}
\end{figure*}

\begin{figure*}[hp!]
	\centering
	\ContinuedFloat
	\plotone{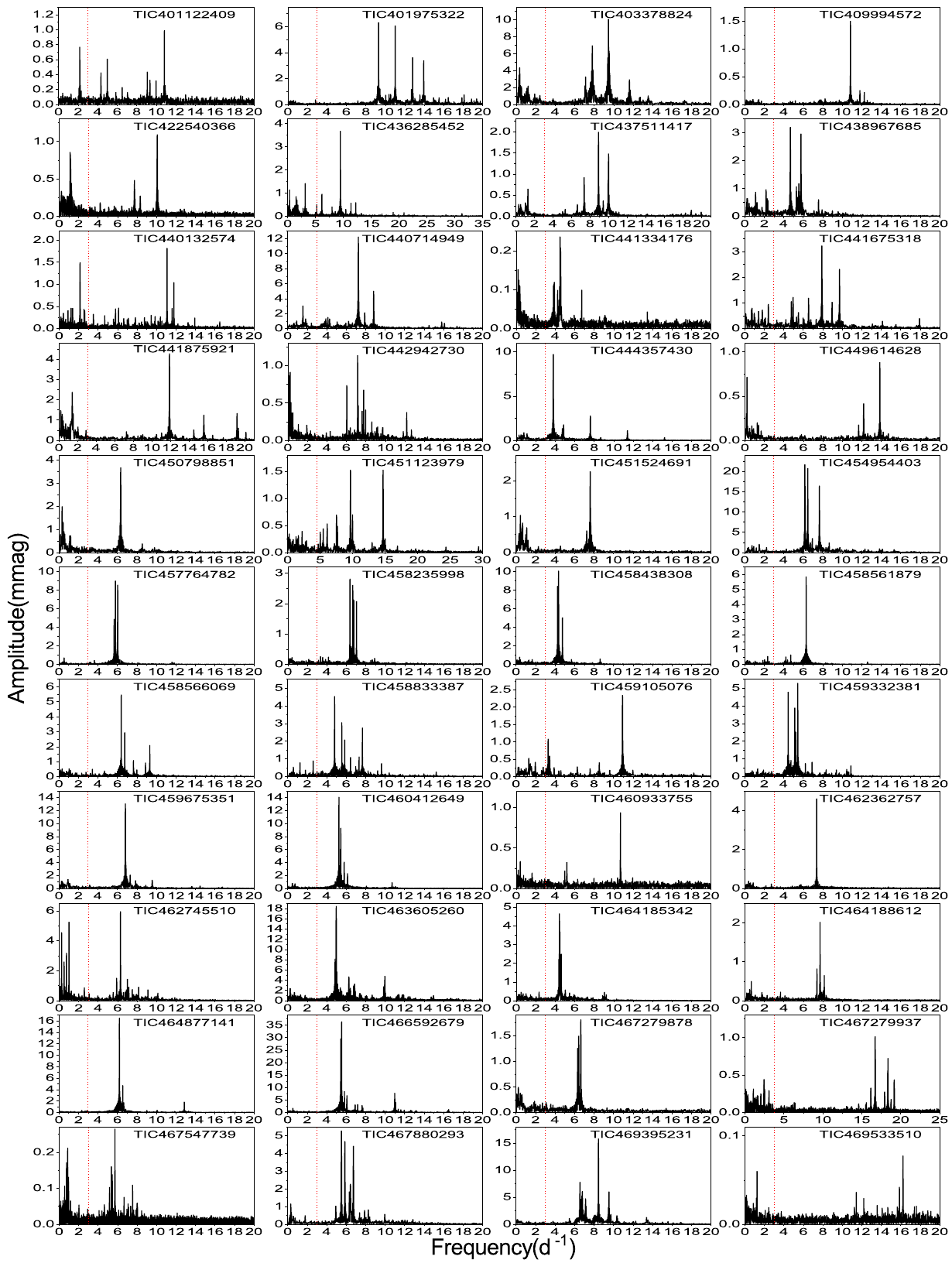}
	\caption[]{(Continued.)}
\end{figure*}

\begin{figure*}[ht!]
	\plotone{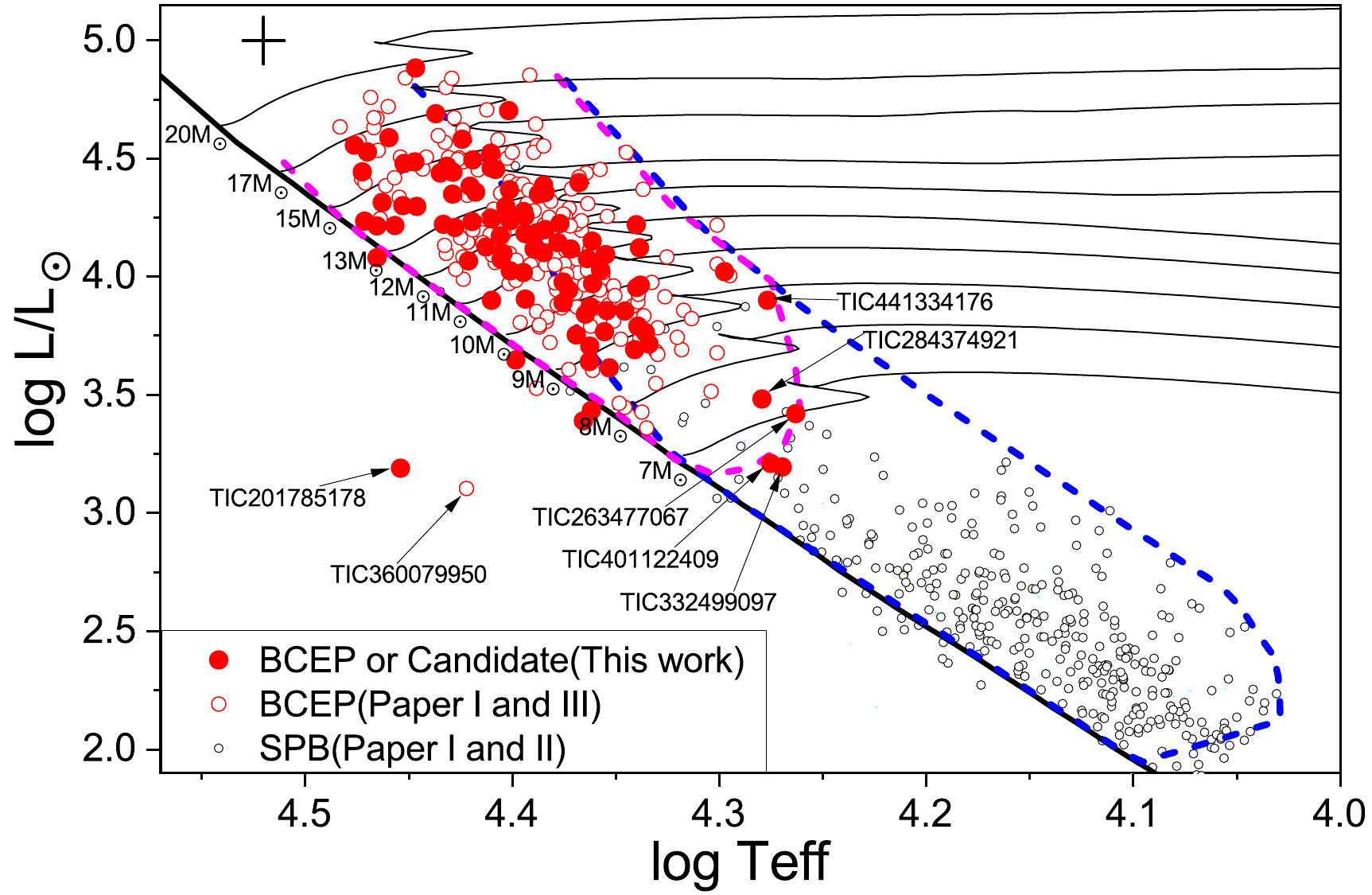}
	\caption{The Hertzsprung--Russell (H--R) diagram of BCEP stars. In this diagram, confirmed BCEP stars and candidates from this work are marked as filled red circles. For comparative analysis, previously published BCEP and SPB stars from Paper I--III are also plotted as open red circles and black circles, respectively. Theoretical stellar evolution frameworks are shown, Black solid curves: Zero-age main sequence (ZAMS) and evolutionary tracks for masses 7--20 $M_{\odot}$ at solar metallicity (Z = 0.02); Blue/magenta dashed curves: the theoretical instability strips for SPB/BCEP stars at Z = 0.02, modeling low-degree pulsation modes $l \leq$ 3 \citep{2007CoAst.151...48M}; Black cross: a standard error box.
		\label{fig:L-T}}
\end{figure*}

\section{Data and Method}
\subsection{TESS and Gaia Data}

TESS, a NASA mission led by MIT, conducts an all-sky survey to detect transiting exoplanets. {TESS employs four identical cameras, each with a 24$^\circ$$\times$24$^\circ$ field of view, which collectively form 24$^\circ$$\times$96$^\circ$ sky strips designated as ``sectors''. For each sector, TESS acquires full-frame images (FFI) at a cadence of 30 minutes or shorter, while simultaneously collecting short-cadence (SC) data of 2-minute or shorter for approximately 20,000 pre-selected targets. The FFI data is selectively processed for brighter stars, and all short-cadence observations are systematically processed through the Science Processing Operations Center (SPOC). The stellar catalog includes nearly one million targets with short-cadence photometry, while the SPOC-processed FFI (FFI-SPOC) data contains more than five times this quantity. Beyond its primary objective of exoplanet detection, TESS's high-precision, large-volume photometric datasets have been proven invaluable for identifying and characterizing bright variable stars (e.g., \citet{2020MNRAS.493.5871B, 2022ApJS..259..50S}).

An analysis of BCEP star characteristics reveals that these objects rarely exhibit pulsation frequencies exceeding 20 $d^{-1}$ (Papers I and III). This observational property makes both the FFI data and the SC data from TESS suitable for their study, since even the FFI data at the maximum 30-minute cadence (corresponding to a Nyquist frequency of 24 $d^{-1}$) is sufficient to capture the highest expected pulsation frequencies.

This study analyzed light curve data obtained from the Mikulski Archive for Space Telescopes (MAST), comprising TESS-SC data for Sectors 1–69 and FFI-SPOC data for Sectors 1–55. Due to the low likelihood of detecting BCEP stars in the FFI-SPOC data, the analysis of Sectors 56–69 is not included in the current work and is scheduled for the next phase. We specifically utilized the Pre-search Data Conditioned Simple Aperture Photometry (PDCSAP) flux measurements \citep{2010ApJ...713L..87J} due to their superior corrections for systematic instrumental variations and their effectiveness in mitigating scattered light contamination compared to standard aperture photometry products. Following the methodology outlined in \citet{2021AJ....161...46S,2021MNRAS.505.6166S}, our pipeline processes the data through flux-to-magnitude conversion and subsequent mean magnitude subtraction, effectively generating light curves that are equivalent to differential photometry.

The Gaia satellite, launched by the European Space Agency (ESA) on December 19, 2013, has collected exceptionally high-precision astrometric data for nearly two billion stars \citep{2016A&A...595A...1G, 2018A&A...616A...1G, 2021A&A...649A...1G}. Its parallax measurements serve as a cornerstone for determining stellar luminosities, providing an independent methodology for investigating variable stars (e.g., \citet{2019MNRAS.485.2380M}) and validating the reliability of astrophysical studies (e.g., \citet{2021PASP..133e4201S}). The stellar atmospheric parameters in Gaia Data Release 3 (DR3) are derived through multiple methodologies that utilize data from the Blue and Red Photometer (BP/RP) low-resolution spectra and Radial Velocity Spectrometer (RVS) observations. Given the high temperatures of our targets, we will employ the stellar atmospheric parameters derived from the Extended Stellar Parametrizer for Hot Stars (ESP-HS) pipeline in this study.

\begin{figure*}[ht!]
	\plotone{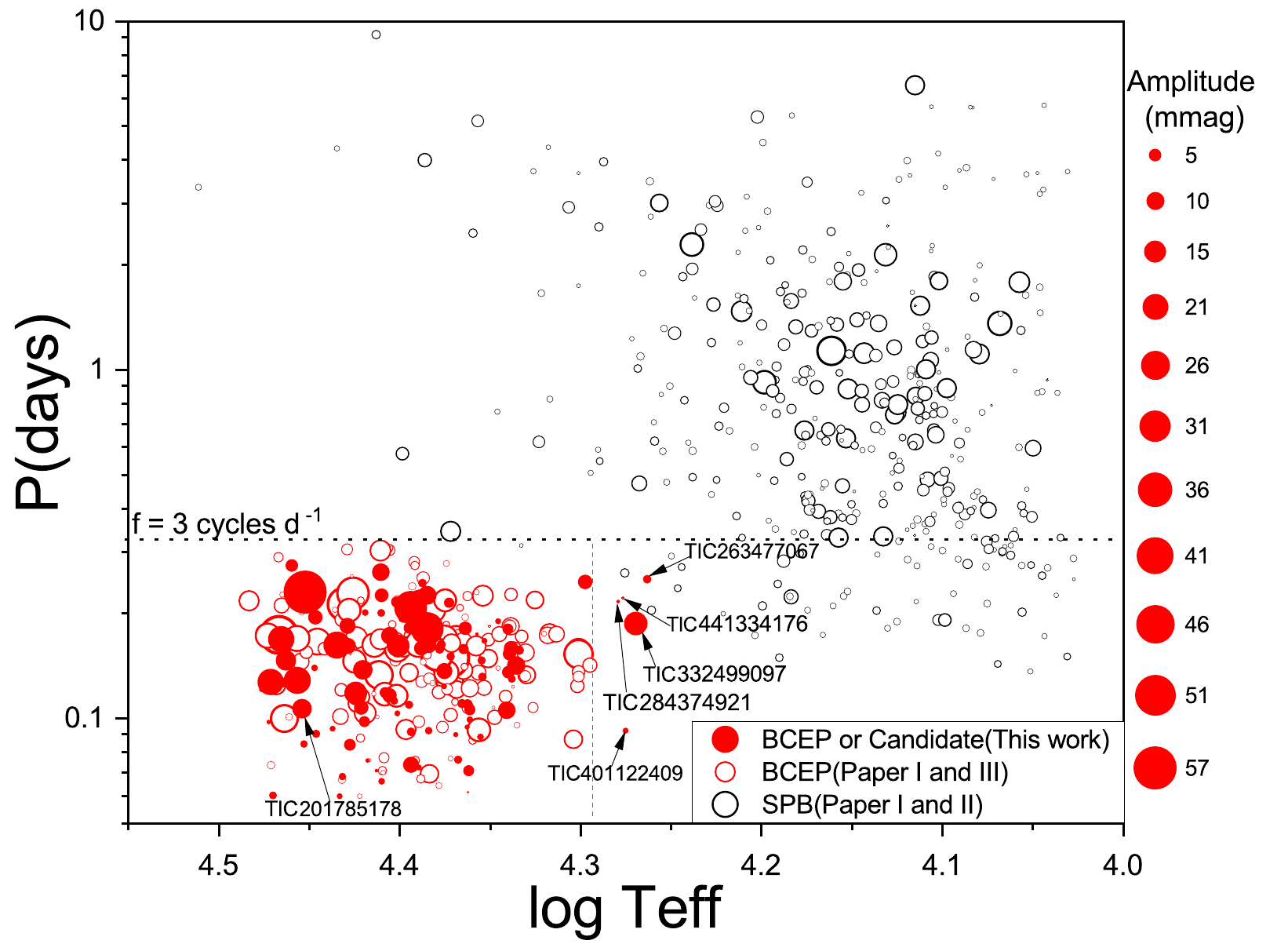}
	\caption{The effective temperature and the dominant pulsating period relation diagram of these BCEP stars and candidates. Similar symbols to those in Figure \ref{fig:L-T} are used, but the size of the circles denotes their pulsation amplitude at the dominant frequency.
		\label{fig:T-P}}
\end{figure*}

\begin{figure*}[ht!]
	\plotone{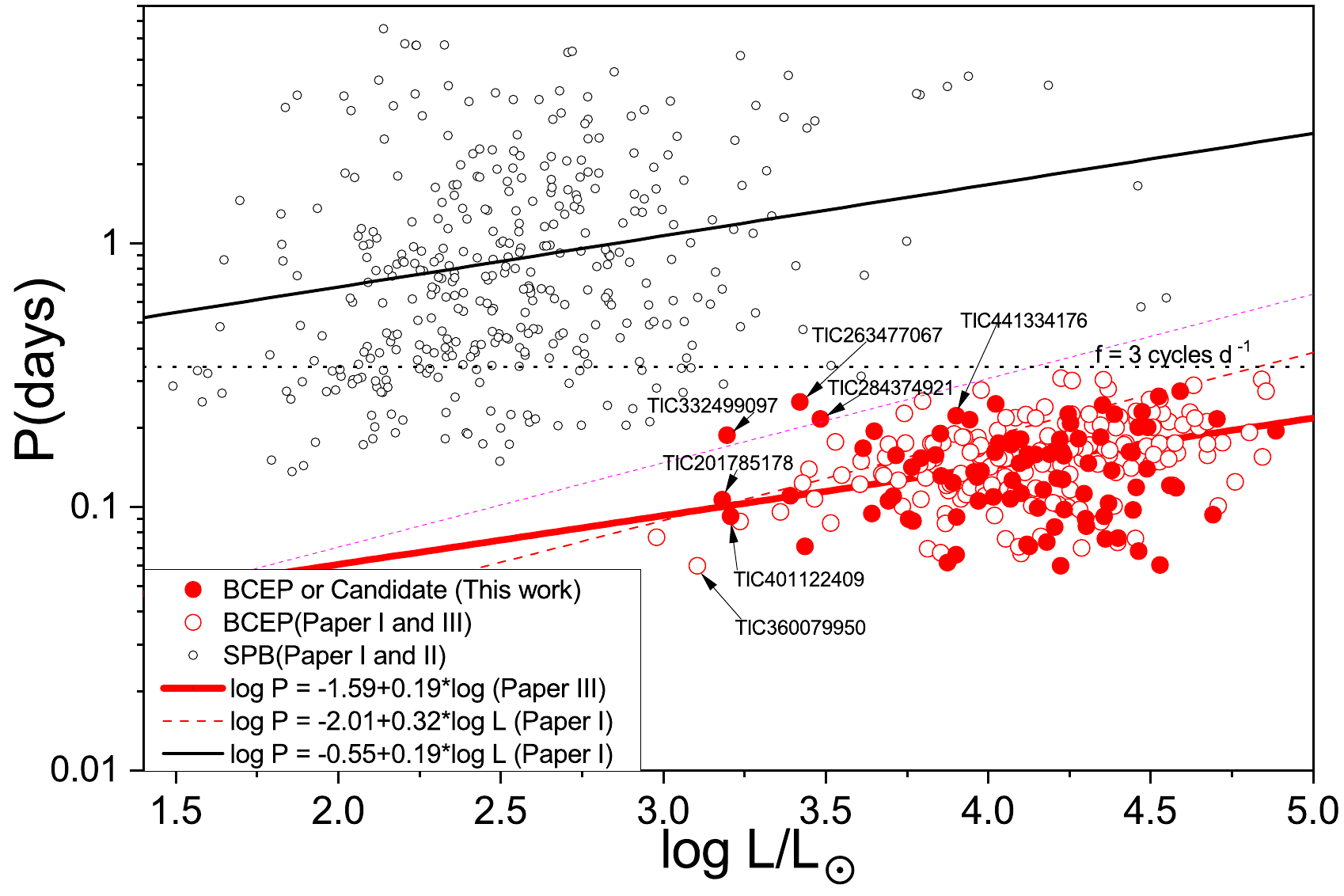}
	\caption{The luminosity and the dominant pulsating period relation diagram of these BCEP stars and candidates. Symbols are similar to those in Figure \ref{fig:L-P}. The red dashed line and the black solid line represent the relationship derived by Paper I for BCEP and SPB stars, respectively. The red solid line represents the new relation derived in Paper III.
		\label{fig:L-P}}
\end{figure*}

\subsection{Classification}

Building on our previous TESS-based investigations of BCEP stars (Papers I and III), this study identifies 85 new BCEP stars or candidates through the analysis of TESS-SC data, with 32 originating from Sectors 1–55 and 53 from Sectors 56–69. Additionally, expanding the survey to include TESS FFI-SPOC data of Sectors 1-55 reveals only three new candidates (TIC201785178, TIC263477067, and TIC401122409). Among these 88 sources, TIC386693012 and TIC459105076, previously classified as candidate pulsators by \citet{2020AJ....160...32L}, are now unambiguously confirmed as BCEP stars for the first time. The fundamental parameters and pulsation properties of these BCEP stars or candidates are summarized in Table \ref{tab:1}.

The detailed identification criteria for BCEP stars are outlined in Papers I and III. These criteria primarily include the Fourier spectral features derived from light curve analysis, as well as the positional distributions in Hertzsprung--Russell (H--R) diagrams, the temperature--period (T--P) diagrams, and the luminosity--period (L--P) diagrams. Except for six objects, which are temporarily classified as candidates for BCEP stars (denoted as "BCEP?") due to their partial fulfillment of the identification criteria, all other targets are ultimately identified as BCEP stars and marked as "BCEP" in column (10) of Table \ref{tab:1}.

Given the angular resolution of TESS (\(21^{\prime\prime}\) per pixel), we conducted a systematic assessment of contamination for all targets by examining potential flux contributions from neighboring stars within the photometric aperture. Utilizing image data from SIMBAD and high-precision astrometry from Gaia, we cross-matched all sources within a 1-arcminute diameter centered on each target, and then .pdfd the capital letter "C" in column 10 of Table \ref{tab:1} to indicate the targets that may be contaminated.

We also employed the TESS-Localize software \citep{2023AJ....165..141H} to determine the spatial origin of the light variations. This tool utilizes observed frequencies as input parameters to compute optimal signal source locations within the specified Target Pixel File (TPF). To evaluate the fitting quality, the software provides two statistical metrics: (1) relative likelihood, which quantifies the probability distribution of potential source stars matching the fitted location; and (2) p-values, which test the hypothesis that each star's position is consistent with the derived signal origin. In this work, a signal is considered to originate from our target source if it has the highest relative likelihood ($>$60\%) and the highest p-value ($>$0.1) among all sources within its TPF. As a result, we found that the prominent signals for nearly all targets originated from their intended stars, with two exceptions. First, we excluded one target (TIC 460689513, it should be a B-type star contaminated by a Delta Scuti variable) originally classified as a BCEP star under other criteria. Second, we confirmed that the apparent variability detected in TIC 464732837 is due to contamination from the nearby pulsating star TIC 464732844.

\section{Results}
\subsection{Stellar Parameters}
Table \ref{tab:1} presents the astrophysical parameters of 88 BCEP stars and candidates analyzed in this study. Column (2) lists the Gaia DR3 parallaxes $\pi$ with typical uncertainties of 0.05 mas. The visual magnitudes $m_{V}$ in Column (3) were systematically derived from the TESS Input Catalog (TIC; \citet{2018AJ....156..102S}) or from SIMBAD \citep{2000A&AS..143....9W}.

The bolometric correction $BC$ in Column (4) was derived from the hybrid local thermodynamic equilibrium/non-local thermodynamic equilibrium (LTE/NLTE) grid of \citet{2020MNRAS.495.2738P}, which provide temperature-dependent calibrations for the Johnson V passband. As demonstrated in our previous work (Papers I and II), the Gaia ESP-HS spectroscopic catalog offers reliable temperature determinations for hot stars, with systematic uncertainties typically below 10\% across the parameter space of BCEP stars. Consequently, both the effective temperatures $T_{eff}$ and interstellar extinction values $A_{V}$ listed in Columns (6) and (5) of Table \ref{tab:1}, respectively, were adopted from the Gaia ESP-HS catalog, applying significant digit selection to ensure their physical relevance.

The stellar luminosities in Column (7) were calculated using the following formulas:
\begin{equation}
	log(L_/L_{\odot})=0.4\times{(4.74-M_{V}-BC)}
\end{equation}
where the absolute magnitude $M_{V}$ is determined by:
\begin{equation}
	M_{V}=m_{V}-5\times{log(1000/\pi)}+5-A_{V},
\end{equation}
Error propagation analysis, which considers typical parameter uncertainties ( $\sigma_{\pi}$ = 0.05 mas, $\sigma_{A_{V}}$ = 0.10 mag, $\sigma_{BC}$ = 0.02 mag, $\sigma_{m_{V}}$ = 0.01,mag), yields a mean luminosity uncertainty of $\sigma_{log L}$ $\approx$ 0.1 dex. This result is consistent with the precision levels reported in recent studies of massive pulsators \citep{2020MNRAS.493.5871B}.

The 88 targets exhibit the visual magnitudes spanning 8.0--12.0 mag, the parallaxes between 0.11 and 1.74 mas, the effective temperatures of 18,000--30,000 K, and the bolometric luminosities ranging from 1,500 to 38,000 $L_\odot$. These parameter distributions are consistent with those reported in Paper III.

\subsection{Pulsation Properties}

The Fourier spectra of the light curves for these BCEP stars and candidates were processed using the Period04 software \citep{2005CoAst.146...53L}, with the resulting frequency analyses presented in Figure \ref{fig:FS}. The signal-to-noise ratio (S/N), defined as the ratio of the pulsation amplitude to the local noise floor (calculated within a 1 $d^{-1}$ frequency bin), serves as the primary criterion for assessing the statistical significance of detected frequencies. Following the methodology of \citet{2021AcA....71..113B}, we adopt a detection threshold of S/N $\geq$ 5.4 for SC data and S/N $\geq$ 5.0 for FFI data.

Adopting the period range of p-mode (P $<$ 0.3 days; \citealt{2010aste.book.....A, 2005ApJS..158..193S}), we empirically define 3 $d^{-1}$ as an operational threshold separating low-frequency and high-frequency domains in the Fourier spectrum. As discussed in Paper III, frequencies above this threshold are unlikely to originate from binarity or rotational modulation, as their corresponding equatorial rotation velocities will be greater than 600 $km \cdot s^{-1}$.

Columns (8)--(9) of Table \ref{tab:1} present the pulsation periods (P) and amplitudes (A) corresponding to the dominant frequencies, with uncertainties in the last decimal place provided in parentheses following the methodology of \citet{1999DSSN...13...28M}. The data reveal pulsational characteristics spanning P=0.06--0.27 days and A =0.1--56 mmag in the TESS band, consistent with typical BCEP variables reported in Paper III.

Figure \ref{fig:L-T}--\ref{fig:L-P} show the positional distributions of these BCEP stars and candidates in H--R diagram, T--P diagram, and L--P diagram. In the H-R diagram, we present the theoretical Zero Age Main Sequence (ZAMS) and the evolutionary tracks of stars with different masses (black solid lines) created using the stellar evolution code Modules for Experiments in Stellar Astrophysics \citep{2011ApJS..192....3P, 2013ApJS..208....4P, 2015ApJS..220...15P, 2018ApJS..234...34P, 2019ApJS..243...10P}, as well as the theoretical unstable regions of SPB and BCEP stars (blue and magenta  dashed lines) given by \citet{2007CoAst.151...48M}.

The H--R diagram (Figure \ref{fig:L-T}) reveals that, with the exception of TIC201785178, all targets occupy the main-sequence region coinciding with the theoretical instability strip of BCEP stars. TIC201785178 displays a characteristic pulsational spectrum of BCEP stars, with a Gaia parallax of 1.59 mas that is the second-highest value among the samples in both this work and Paper III. Notably, SIMBAD classifies this object as a double or multiple star system. It is likely a BCEP star, and its lower luminosity may be caused by unreliable parallax measurements due to the orbital motion of binary or multiple system, but it may also be a subdwarf star in a binary system. The situation of TIC201785178 is very similar to TIC360079950 discussed in Paper III.

The T--P diagram (Figure \ref{fig:T-P}) demonstrates that all targets except five objects (TIC263477067, TIC441334176, TIC332499097, TIC284374921, and TIC401122409) are co-located with the BCEP stars in the same parameter space as those reported in Paper III. 
TIC263477067, TIC284374921, TIC332499097, and TIC401122409 are located in the low-mass (red) end of the theoretical pulsational instability strip, which is usually sparsely populated in observational surveys (e.g., \citet{2005ApJS..158..193S}). Furthermore, our analysis reveals that their light curves yield a limited number ($<$ 10) of significant pulsation frequencies, likely attributable to restricted mode excitation near the theoretical instability strip boundary. Therefore, although their distribution in the H-R, T-P, or L-P diagrams is different from other BCEP stars, they are still suspected to be BCEP stars, and marked as candidates. They will be excellent samples for studying the red end of theoretical instability strip. Unlike the other four targets, TIC 441334176 exhibits more than ten significant pulsation frequencies, and its location on the H-R diagram preliminarily indicates that it is an 8–9 $M_{\odot}$ star near the main-sequence turnoff. In comparison, the other four targets are probable main-sequence stars of below 7.5 $M_{\odot}$ located nearer to the red edge.

The L-P diagram (Figure \ref{fig:L-P}) reveals that most these new objects are consistent with the empirical relationship established in Paper III. However, TIC263477067, TIC332499097, and TIC284374921 display significant deviations from this relationship.

\begin{figure*}[ht!]
	\plotone{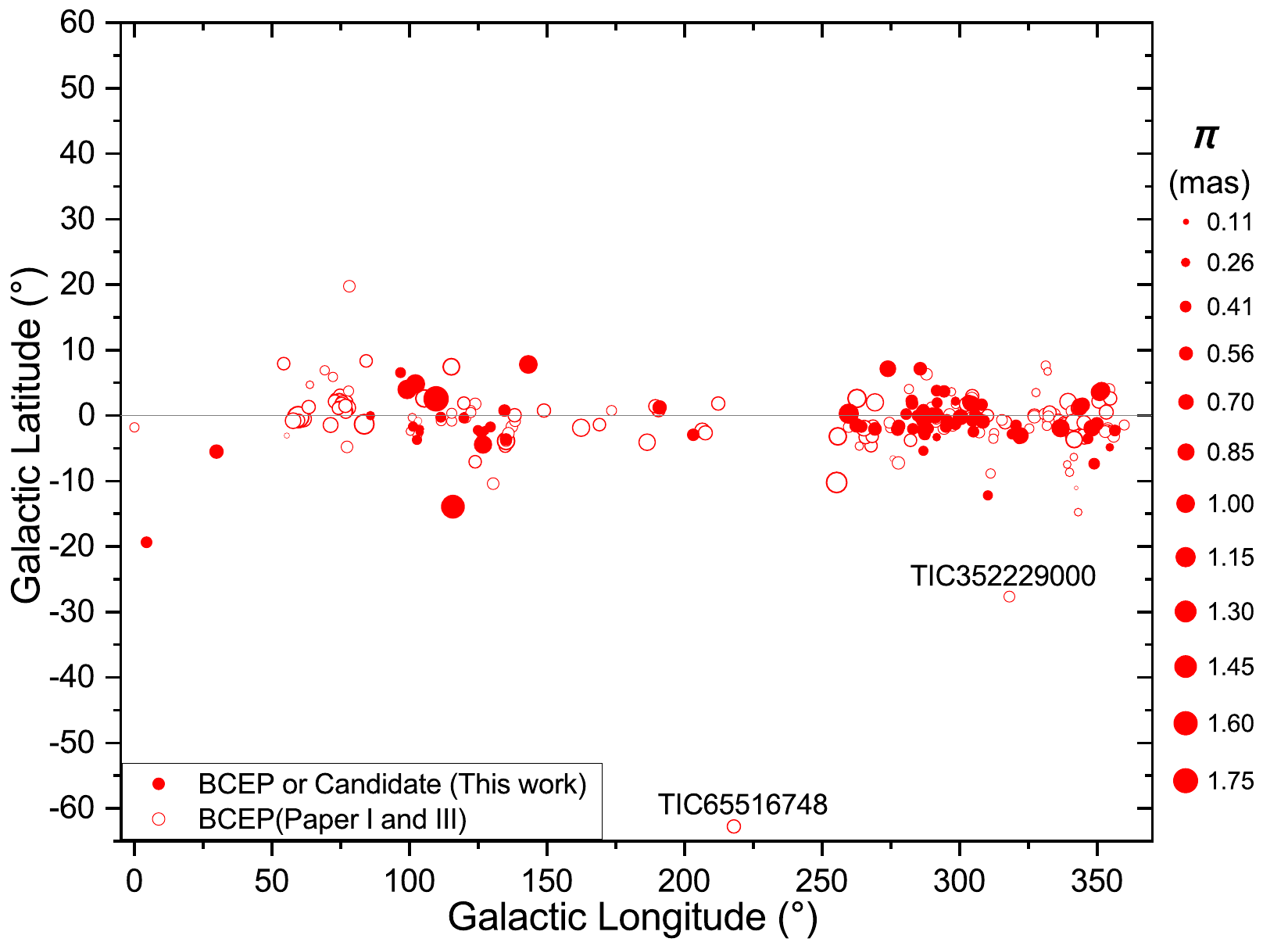}
	\caption{The Distribution of BCEP stars according to Galactic longitude and latitude. Similar symbols to these BCEP stars in Figure \ref{fig:L-T} are used, but the size of the circles denotes their parallaxes $\pi$.
		\label{fig:b-I}}
\end{figure*}

\subsection{The Spatial Distribution of $\beta$ Cephei Stars}

The galactic spatial distribution of BCEP stars from this study (solid circles), as well as from Papers I and III (open circles), is illustrated in Figure \ref{fig:b-I}. The sizes of the symbols correspond to the Gaia parallax $\pi$, with larger symbols indicating shorter distances. These samples encompasses heliocentric distances ranging from 0.6 to 8 kpc, with a predominant concentration in the Galactic disk (Galactic latitude $|b| < 20^\circ$).

Notable outliers, TIC352229000 and TIC65516748, located at high Galactic latitudes, likely formed in the Galactic halo or represent runaway stars ejected from the disk \citep{2001A&A...378..907R}. Gaia DR3 astrometry reveals exceptionally high proper motions for both targets: TIC352229000 exhibits $(\mu_\alpha, \mu_\delta) = (16.48, -18.72) \, \mathrm{mas\,yr^{-1}}$, while TIC65516748 displays $(\mu_\alpha, \mu_\delta) = (-6.78, -7.80) \, \mathrm{mas\,yr^{-1}}$. These values significantly exceed the median proper motion of Galactic disk stars ($\sim$1--5 mas yr$^{-1}$) \citep{2021A&A...649A...1G}, translating to transverse velocities of $\sim$317 km s$^{-1}$ and 85 km s$^{-1}$, respectively. This is a relatively large value, we classify both objects as runaway stars, likely ejected through either dynamical interactions in dense stellar environments or binary supernova recoil mechanisms.

\begin{figure*}[ht!]
	\plotone{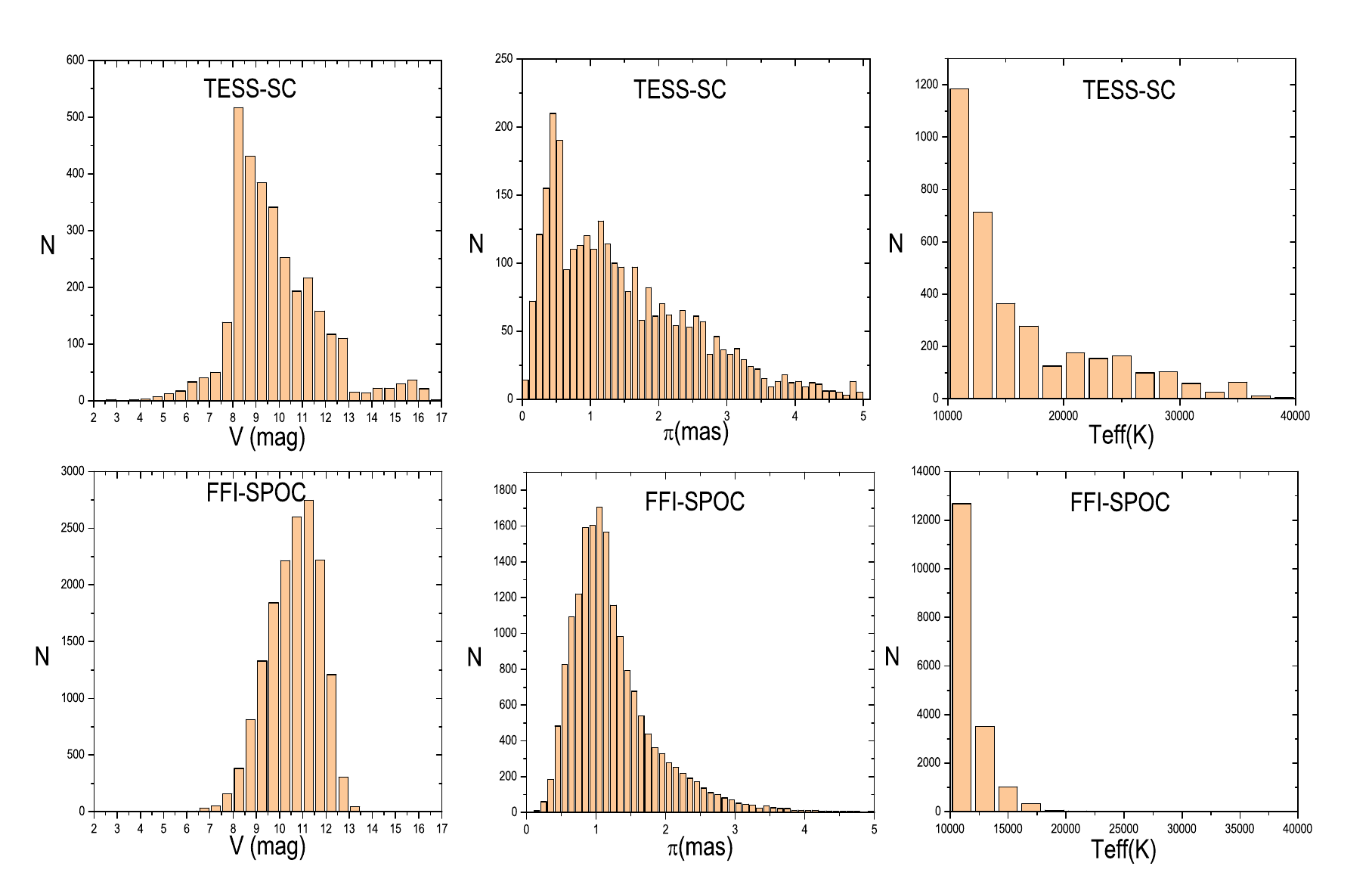}
	\caption{Distribution of the parallax, visual magnitude, and temperature for high-temperature stars ($T > 10,000 K$) with TESS-SC and FFI-SPOC data.
		\label{fig:distribution}}
\end{figure*}

\subsection{Detection Constraints of TESS}

We compared the parallax, visual magnitude, and temperature distribution of high-temperature stars ($T > 10,000 K$ provided by GAIA ESP-HS) with TESS-SC and FFI-SPOC data from Sector 1 to 55. As illustrated in Figure \ref{fig:distribution}, the peak visual magnitudes for TESS-SC and FFI-SPOC targets are centered at 8 mag and 11 mag, respectively, with corresponding peak parallax values of 0.5 mas and 1.0 mas. This correlation between visual magnitude and parallax suggests that FFI-SPOC predominantly captures these lower luminosity stars compared to TESS-SC.

Assuming these stars are main-sequence stars, the lower luminosity of FFI-SPOC targets further implies systematically cooler temperatures relative to their TESS-SC counterparts, a trend quantitatively supported by the temperature histograms in the right panels of Figure \ref{fig:distribution}. Among stars with TESS-SC data, approximately 30\% exhibit temperatures exceeding 18,000 K, whereas only 0.7\% of stars with FFI-SPOC data have temperatures exceeding 18000K. Therefore, only 3 candidates were identified within the FFI-SPOC data in this work. Notably, our companion study (Paper II), which analyzed hot star populations cataloged by LAMOST spectroscopic surveys, did not detect any confirmed BCEP stars or candidates.

The angular resolution of TESS (\(21^{\prime\prime}\) per pixel) imposes significant challenges in resolving targets within densely populated regions of the Galactic disk, particularly toward the Galactic center. In such crowded fields, distant BCEP stars (located at heliocentric distances \(>5\)~kpc) are prone to blending with neighboring stars within the photometric aperture. This contamination artificially inflates flux measurements, suppresses detectable pulsation amplitudes, and complicates the extraction of clean light curves. For example, in the Galactic plane near the bulge (\(|\ell| < 30^\circ\), \(|b| < 10^\circ\)), the high stellar density (exceeding \(10^3\)~stars~deg\(^{-2}\); \citep{2021A&A...649A...1G}) exacerbates these issues, leading to potential misclassifications or missed detections of genuine BCEP stars. This effect partially explains the scarcity of BCEP candidates in FFI-SPOC data and the absence of BCEP stars in Paper~II's LAMOST sample.

However, these conclusions are primarily a result of observational selection effects, and they may change with evolving observing strategies and the pre-selected targets of TESS. For instance, the strategy in TESS Sectors 56–69 targeted numerous new massive stars, which is the direct reason we were able to identify 53 additional targets from these data.

\section{Discussion and conclusion}
Through the integration of TESS photometric data with Gaia spectral and astrometric observations, we have identified 82 new confirmed BCEP stars and 6 candidates from TESS-SC and FFI-SPOC data. These discoveries indicate the parameter space of BCEP stars to visual magnitudes of 8.0--12.0 mag, parallaxes of 0.11--1.74 mas, effective temperatures of 18,000--30,000 K, and luminosities spanning 1,500--38,000 $L_\odot$, consistent with the characteristics of BCEP stars outlined in Paper III.

In the H-R diagram, the anomalous object TIC201785178 attributed to potential binarity or multiplicity, mirrors the case of TIC360079950 in Paper III. Such systems require follow-up radial velocity monitoring to disentangle orbital and pulsational signals, as unresolved companions may distort parallax-based luminosity estimates. Similarly, TIC263477067, TIC284374921, TIC332499097, and TIC401122409 are located near the red boundary of the instability strip in the H-R diagram, and their inhibited pulsation mode excitation suggests that they are excellent samples for testing the red boundary of the theoretical pulsation model for BCEP stars.

The spatial distribution of BCEP stars, as shown in Figure 5, demonstrates a significant concentration within the Galactic disk ($|b| < 20^\circ$). This pattern aligns with their massive, short-lived characteristics and their origins in star-forming regions. The outliers TIC352229000 and TIC65516748, located at high Galactic latitudes, likely represent runaway stars that have been ejected from the disk \citep{2001A&A...378..907R} or remnants of disrupted dwarf galaxies. These stars provide valuable insights into the dynamical processes occurring in the halo.

The comparative analysis of the TESS-SC and FFI-SPOC datasets reveals significant observational biases in the detection of BCEP stars. TESS-SC targets predominantly capture brighter, higher-mass stars, but FFI-SPOC data are skewed toward fainter, lower-luminosity objects. This luminosity disparity corresponds to a systematic temperature difference--only 0.7\% of FFI-SPOC targets exceed 18,000~K, compared to 30\% in the TESS-SC data. Consequently, FFI-SPOC observations yielded only three BCEP candidates, highlighting the challenges of identifying high-temperature pulsators at fainter magnitudes. The scarcity of BCEP stars in the FFI-SPOC data may also reflect limitations in photometric precision for low-amplitude pulsations or contamination from blended sources due to TESS's coarse angular resolution of \(21^{\prime\prime}\) per pixel. Future missions with higher spatial resolution and extended time-series coverage will be essential for disentangling blended sources and exploring the distant, high-density regions of the Galactic disk where massive pulsators may remain concealed.

These results underscore the significance of high-precision, uninterrupted space photometry from TESS in identifying low-amplitude bright pulsators, as well as the value of Gaia's astrometric precision in validating astrophysical parameters. As TESS data continue to be updated and only a subset has been analyzed thus far, the present study should not be regarded as complete. Future work will focus on broadening the search scope and performing a systematic reclassification of BCEP stars identified by other teams, with the overarching aim of establishing a consistent characterization framework for the entire BCEP class. In subsequent studies, priority should also be given to the following: high-resolution spectroscopy to confirm binarity in outliers (e.g., TIC 201785178) and to measure rotational velocities; and extended pulsation mode analysis for samples near the boundaries of the instability strip.

\clearpage

%\onecolumn
\begin{longtable}{llllllllllll}
	
	\caption{\label{tab:1} The Catalog of BCEP stars and candidates newly identified from TESS and Gaia data.}\\
	\hline
	TESS ID   & $\pi$   & V       & $BC$    & $A_{V}$ & $Teff$   & $log(L_/L_{\odot})$    & Period     & Amplitude & Comments & Flag \\
	& ($mas$) & ($Mag$) & ($Mag$) & ($Mag$) & ($K$)    &                        & ($days$)   & ($mmag$)  &          &      \\
	\hline
	\endfirsthead
	
	\caption{\label{tab:1}(Continued)}\\
	\hline
	TESS ID   & $\pi$   & V       & $BC$    & $A_{V}$ & $Teff$   & $log(L_/L_{\odot})$    & Period     & Amplitude & Comments & Flag \\
	& ($mas$) & ($Mag$) & ($Mag$) & ($Mag$) & ($K$)    &                        & ($days$)   & ($mmag$)  &          &      \\
	\hline \endhead
	\hline
	\multicolumn{6}{r}{\textsl{(Continued)}}\\
	\endfoot
	\hline
	\endlastfoot
	
	12794506  & 1.74 & 9.90  & -2.79 & 4.33 & 28400(400) & 4.30 & 0.084451671(1) & 1.42 (1) & BCEP     &   \\
	13773736  & 0.45 & 13.45 & -2.48 & 3.63 & 25000(500) & 3.65 & 0.194130664(1) & 2.65 (4) & BCEP(C)  &   \\
	31557506  & 0.38 & 8.89  & -1.92 & 0.16 & 19800(100) & 4.02 & 0.246114383(3) & 6.05 (1) & BCEP     &   \\
	33678125  & 0.38 & 9.97  & -2.81 & 0.85 & 28600(500) & 4.21 & 0.128415381(1) & 21.82(1) & BCEP     &   \\
	35909965  & 0.74 & 9.45  & -2.30 & 0.23 & 23200(400) & 3.39 & 0.110331215(1) & 2.85 (1) & BCEP     &   \\
	40066259  & 0.88 & 9.36  & -2.46 & 1.92 & 24800(500) & 4.01 & 0.109273377(1) & 1.84 (1) & BCEP     &   \\
	42534267  & 0.94 & 8.35  & -2.36 & 0.84 & 23700(300) & 3.89 & 0.123628955(1) & 0.81 (1) & BCEP(C)  &   \\
	43296080  & 0.85 & 8.73  & -2.55 & 0.85 & 25700(500) & 3.90 & 0.065957204(1) & 1.32 (1) & BCEP     &   \\
	43460530  & 0.41 & 10.43 & -2.76 & 3.20 & 28000(500) & 4.88 & 0.194801805(1) & 5.84 (1) & BCEP(C)  &   \\
	52696829  & 0.31 & 10.80 & -2.55 & 2.12 & 25700(500) & 4.46 & 0.200818875(3) & 1.95 (2) & BCEP(C)  &   \\
	54371344  & 0.36 & 10.70 & -2.59 & 2.06 & 26200(600) & 4.36 & 0.075740829(1) & 0.49 (1) & BCEP     &   \\
	63969423  & 0.16 & 11.87 & -2.31 & 1.83 & 23300(400) & 4.40 & 0.076134460(1) & 1.82 (1) & BCEP     &   \\
	64077872  & 0.22 & 11.63 & -2.56 & 1.36 & 25900(500) & 4.13 & 0.070792469(1) & 0.25 (1) & BCEP(C)  &   \\
	110879461 & 0.38 & 11.69 & -2.90 & 2.54 & 29600(600) & 4.23 & 0.126993380(1) & 20.88(2) & BCEP     &   \\
	120859681 & 0.51 & 8.88  & -2.37 & 0.68 & 23900(200) & 4.14 & 0.158119787(1) & 2.78 (1) & BCEP     &   \\
	141365778 & 0.55 & 10.62 & -2.90 & 2.78 & 29700(400) & 4.44 & 0.097590761(1) & 0.39 (1) & BCEP     &   \\
	180540045 & 1.13 & 8.98  & -2.28 & 0.81 & 23000(400) & 3.43 & 0.070782259(1) & 3.58 (1) & BCEP     &   \\
	196639455 & 0.78 & 9.48  & -2.28 & 1.19 & 23000(500) & 3.71 & 0.109794376(1) & 3.39 (1) & BCEP(C)  &   \\
	201785178 & 1.59 & 10.96 & -2.80 & 2.37 & 28400(2100)& 3.18 & 0.106536298(1) & 12.14(2) & BCEP?(C) &   \\
	217754401 & 0.33 & 10.01 & -2.50 & 2.12 & 25200(500) & 4.70 & 0.215771750(3) & 3.53 (1) & BCEP(C)  &   \\
	263477067 & 0.41 & 11.10 & -1.72 & 1.27 & 18300(400) & 3.42 & 0.250900812(1) & 2.18 (2) & BCEP?    &   \\
	268027536 & 0.45 & 9.39  & -2.53 & 0.80 & 25400(300) & 4.17 & 0.116430153(1) & 6.32 (1) & BCEP(C)  &   \\
	279168450 & 0.23 & 11.52 & -2.50 & 2.03 & 25200(500) & 4.37 & 0.103186418(1) & 1.00 (2) & BCEP(C)  &   \\
	284374921 & 0.75 & 9.51  & -1.82 & 1.03 & 19000(100) & 3.48 & 0.216285812(2) & 0.33 (1) & BCEP?(C) &   \\
	305918859 & 0.41 & 9.82  & -2.60 & 1.10 & 26300(500) & 4.23 & 0.097811596(1) & 3.85 (1) & BCEP(C)  &   \\
	315798707 & 0.45 & 8.92  & -2.92 & 0.94 & 29900(300) & 4.56 & 0.120953136(1) & 0.70 (1) & BCEP     &   \\
	319376730 & 0.58 & 9.40  & -2.15 & 0.79 & 21800(300) & 3.79 & 0.153347065(1) & 5.19 (1) & BCEP(C)  &   \\
	320732406 & 0.43 & 10.64 & -2.32 & 1.15 & 23400(400) & 3.75 & 0.089871619(1) & 0.63 (2) & BCEP     &   \\
	322418939 & 0.39 & 9.01  & -2.65 & 0.65 & 26800(400) & 4.44 & 0.161175643(2) & 7.74 (2) & BCEP(C)  &   \\
	330241794 & 0.27 & 9.96  & -2.83 & 1.01 & 28800(300) & 4.59 & 0.274758202(1) & 4.69 (1) & BCEP(C)  &   \\
	332158408 & 0.37 & 9.80  & -2.42 & 1.33 & 24400(300) & 4.35 & 0.244061039(3) & 2.42 (1) & BCEP     &   \\
	332499097 & 0.88 & 10.02 & -1.76 & 1.23 & 18600(300) & 3.19 & 0.187262382(2) & 16.73(2) & BCEP?(C) &   \\
	339561272 & 0.64 & 10.01 & -2.24 & 1.48 & 22700(500) & 3.77 & 0.088337148(1) & 1.35 (2) & BCEP(C)  &   \\
	339570192 & 0.64 & 10.20 & -2.28 & 1.32 & 23100(600) & 3.64 & 0.094294761(1) & 0.68 (1) & BCEP(C)  &   \\
	341262179 & 0.62 & 9.26  & -2.86 & 1.15 & 29200(200) & 4.22 & 0.168535031(1) & 20.58(1) & BCEP(C)  &   \\
	344542087 & 0.33 & 10.70 & -2.36 & 1.78 & 23800(500) & 4.23 & 0.162795972(1) & 2.82 (1) & BCEP     &   \\
	372627092 & 0.34 & 10.70 & -2.41 & 2.18 & 24300(400) & 4.39 & 0.225308235(1) & 11.40(1) & BCEP     &   \\
	377768377 & 0.27 & 9.72  & -2.16 & 0.57 & 21900(400) & 4.22 & 0.180602029(1) & 3.80 (1) & BCEP     &   \\
	377895204 & 0.33 & 10.68 & -2.54 & 2.14 & 25600(500) & 4.45 & 0.118484236(1) & 3.67 (1) & BCEP(C)  &   \\
	381259601 & 0.38 & 9.65  & -2.24 & 0.80 & 22600(400) & 4.09 & 0.146601333(1) & 1.87 (1) & BCEP     &   \\
	383090276 & 0.25 & 11.32 & -2.89 & 1.96 & 29500(700) & 4.53 & 0.060149613(1) & 2.09 (3) & BCEP(C)  &   \\
	386693012 & 0.54 & 8.62  & -2.46 & 0.53 & 24800(400) & 4.18 & 0.073546294(1) & 7.58 (1) & BCEP     &(1)\\
	391516015 & 0.40 & 9.62  & -2.46 & 1.10 & 24800(500) & 4.28 & 0.181821400(1) & 4.14 (1) & BCEP(C)  &   \\
	393346713 & 0.43 & 10.52 & -2.52 & 1.66 & 25400(500) & 4.10 & 0.112754739(1) & 0.62 (1) & BCEP     &   \\
	401122409 & 0.51 & 12.03 & -1.79 & 2.03 & 18800(300) & 3.21 & 0.092289677(1) & 0.98 (3) & BCEP?    &   \\
	401975322 & 0.29 & 10.27 & -2.61 & 0.42 & 26400(500) & 4.07 & 0.107368784(1) & 6.31 (1) & BCEP     &   \\
	403378824 & 0.54 & 10.21 & -2.16 & 1.20 & 21900(600) & 3.69 & 0.105572076(1) & 8.90 (1) & BCEP(C)  &   \\
	409994572 & 1.10 & 7.99  & -2.40 & 1.95 & 24200(400) & 4.36 & 0.092277609(1) & 1.50 (1) & BCEP     &   \\
	422540366 & 0.26 & 11.11 & -2.28 & 1.54 & 23000(400) & 4.15 & 0.099244614(1) & 1.09 (2) & BCEP(C)  &   \\
	436285452 & 0.50 & 9.43  & -2.28 & 0.85 & 23000(400) & 3.97 & 0.105642375(1) & 3.67 (1) & BCEP(C)  &   \\
	437511417 & 0.54 & 9.28  & -2.51 & 1.42 & 25300(400) & 4.29 & 0.112221025(1) & 1.99 (1) & BCEP     &   \\
	438967685 & 0.71 & 9.88  & -2.34 & 1.92 & 23600(400) & 3.94 & 0.214345642(1) & 3.40 (1) & BCEP(C)  &   \\
	440132574 & 0.47 & 10.99 & -2.75 & 2.60 & 27900(600) & 4.30 & 0.090247548(1) & 1.79 (1) & BCEP(C)  &   \\
	440714949 & 0.47 & 10.32 & -2.61 & 2.28 & 26300(700) & 4.38 & 0.137889592(1) & 12.22(1) & BCEP(C)  &   \\
	441334176 & 0.47 & 9.92  & -1.80 & 1.48 & 18900(300) & 3.90 & 0.222130887(1) & 0.19 (1) & BCEP?    &   \\
	441675318 & 0.43 & 10.81 & -2.86 & 1.55 & 29200(300) & 4.07 & 0.126367362(1) & 3.23 (1) & BCEP     &   \\
	441875921 & 0.39 & 9.61  & -2.65 & 0.70 & 26800(500) & 4.20 & 0.084019404(1) & 4.27 (1) & BCEP     &   \\
	442942730 & 0.37 & 10.29 & -2.76 & 1.82 & 28000(400) & 4.49 & 0.139608794(1) & 1.13 (1) & BCEP     &   \\
	444357430 & 0.35 & 10.04 & -2.55 & 1.75 & 25700(500) & 4.52 & 0.262674343(1) & 9.11 (1) & BCEP     &   \\
	449614628 & 0.48 & 9.27  & -2.43 & 0.81 & 24500(400) & 4.12 & 0.072207624(1) & 0.88 (1) & BCEP(C)  &   \\
	450798851 & 0.65 & 9.48  & -2.29 & 1.12 & 23200(400) & 3.84 & 0.157869159(1) & 3.35 (1) & BCEP(C)  &   \\
	451123979 & 0.36 & 9.16  & -2.67 & 0.66 & 27000(200) & 4.46 & 0.068105858(1) & 1.53 (1) & BCEP(C)  &   \\
	451524691 & 0.37 & 10.20 & -2.24 & 0.68 & 22600(300) & 3.85 & 0.131394053(1) & 2.25 (2) & BCEP(C)  &   \\
	454954403 & 0.33 & 10.76 & -2.69 & 2.01 & 27200(1000)& 4.43 & 0.162146735(2) & 21.64(2) & BCEP     &   \\
	457764782 & 0.41 & 10.10 & -2.52 & 1.07 & 25400(600) & 4.07 & 0.172588294(1) & 9.00 (1) & BCEP     &   \\
	458235998 & 0.48 & 9.60  & -2.12 & 0.42 & 21600(400) & 3.71 & 0.156935036(1) & 2.79 (1) & BCEP(C)  &   \\
	458438308 & 0.40 & 10.57 & -2.79 & 2.23 & 28300(500) & 4.47 & 0.230050944(1) & 56.49(5) & BCEP     &   \\
	458561879 & 0.46 & 8.77  & -2.15 & 0.51 & 21800(400) & 4.12 & 0.158703061(1) & 5.85 (1) & BCEP(C)  &   \\
	458566069 & 0.40 & 9.72  & -2.50 & 1.09 & 25200(500) & 4.23 & 0.156554387(1) & 5.46 (1) & BCEP(C)  &   \\
	458833387 & 0.42 & 10.32 & -2.46 & 1.89 & 24800(500) & 4.25 & 0.207987915(1) & 32.94(1) & BCEP     &   \\
	459105076 & 0.97 & 9.59  & -2.46 & 2.07 & 24800(300) & 3.90 & 0.091531988(1) & 2.34 (1) & BCEP(C)  &(1)\\
	459332381 & 0.44 & 9.78  & -2.66 & 1.47 & 26800(400) & 4.34 & 0.184129524(1) & 7.97 (1) & BCEP(C)  &   \\
	459675351 & 0.39 & 9.95  & -2.85 & 1.10 & 29000(300) & 4.31 & 0.146690576(1) & 13.03(1) & BCEP(C)  &   \\
	460412649 & 0.33 & 10.50 & -2.19 & 0.77 & 22200(400) & 3.85 & 0.190225162(1) & 1.01 (1) & BCEP(C)  &   \\
	460933755 & 0.11 & 11.49 & -2.70 & 0.98 & 27300(1200)& 4.69 & 0.093386494(1) & 0.95 (2) & BCEP(C)  &   \\
	462362757 & 0.38 & 11.19 & -2.16 & 2.08 & 21900(500) & 3.95 & 0.135783950(1) & 4.61 (1) & BCEP     &   \\
	462745510 & 0.41 & 9.90  & -2.43 & 1.26 & 24400(500) & 4.19 & 0.158740481(1) & 5.96 (1) & BCEP     &   \\
	463605260 & 0.49 & 10.26 & -2.60 & 2.58 & 26200(800) & 4.49 & 0.200982894(1) & 1.13 (1) & BCEP     &   \\
	464185342 & 0.47 & 9.36  & -2.55 & 1.02 & 25700(400) & 4.25 & 0.225357831(1) & 5.73 (1) & BCEP(C)  &   \\
	464188612 & 0.41 & 9.99  & -2.14 & 1.09 & 21800(500) & 3.97 & 0.129677312(1) & 1.99 (1) & BCEP(C)  &   \\
	464877141 & 0.33 & 10.85 & -2.50 & 1.24 & 25200(600) & 4.02 & 0.161205576(1) & 16.50(2) & BCEP     &   \\
	466592679 & 0.22 & 11.73 & -2.41 & 1.52 & 24300(700) & 4.10 & 0.180720462(1) & 33.10(2) & BCEP(C)  &   \\
	467279878 & 0.41 & 10.28 & -2.33 & 1.56 & 23500(400) & 4.11 & 0.150349062(2) & 1.81 (2) & BCEP(C)  &   \\
	467279937 & 0.44 & 10.22 & -2.68 & 1.58 & 27100(700) & 4.22 & 0.059846347(1) & 1.01 (1) & BCEP(C)  &   \\
	467547739 & 0.28 & 10.42 & -2.25 & 0.77 & 22800(300) & 4.03 & 0.174603422(1) & 0.27 (1) & BCEP     &   \\
	467880293 & 0.43 & 9.33  & -2.29 & 0.66 & 23100(400) & 4.07 & 0.181329693(1) & 5.27 (1) & BCEP     &   \\
	469395231 & 0.22 & 10.56 & -2.63 & 1.37 & 26600(800) & 4.58 & 0.118241552(1) & 15.81(3) & BCEP(C)  &   \\
	469533510 & 1.07 & 8.65  & -2.28 & 1.45 & 23000(500) & 3.87 & 0.061591120(1) & 0.07 (1) & BCEP     &   \\
	
\end{longtable}
{Note. Values in parentheses in Columns 6, 8, and 9 indicate the uncertainties. The capital letter "C" in Column 10 flags targets potentially affected by flux contamination from nearby sources within TESS apertures. Two objects previously listed as candidate of BCEP stars by \citet{2020AJ....160...32L} are annotated with "(1)" in Column 11.\\}

\begin{acknowledgments}
This work is partly supported by the International Cooperation Projects of the National Key R\&D Program (No. 2022YFE0116800), the International Partership Program of Chinese Academy of Sciences (No. 020GJHZ2023030GC), Chinese Natural Science Foundation (No. 12273103), Yunnan Revitalization Talent Support Program, and the basic research project of Yunnan Province (Grant Nos. 202501AS070055 and 202301AT070352). It is also funded by Chinese Academy of Sciences President's International Fellowship Initiative (Grant No. 2025PVA0089).
This work has made use of data from the European Space Agency (ESA) mission Gaia (https://www.cosmos.esa.int/gaia), processed by the Gaia Data Processing and Analysis Consortium (DPAC, https://www.cosmos.esa.int/web/gaia/dpac/consortium). Funding for the DPAC has been provided by national institutions, in particular the institutions participating in the Gaia Multilateral Agreement.
The TESS data presented in this paper were obtained from the Mikulski Archive for Space Telescopes (MAST) at the Space Telescope Science Institute (STScI). STScI is operated by the Association of Universities for Research in Astronomy, Inc. Support to MAST for these data is provided by the NASA Office of Space Science. Funding for the TESS mission is provided by the NASA Explorer Program. We are deeply appreciative of the insightful feedback provided by the anonymous referee, which has immensely improved the clarity and rigor of our paper.
\end{acknowledgments}


%% Include this line if you are using the \.pdfd, \replaced, \deleted


\begin{thebibliography}{99}
	
	\bibitem[Aerts et al.(2010)]{2010aste.book.....A} Aerts, C., Christensen-Dalsgaard, J., \& Kurtz, D.~W.\ 2010, Asteroseismology (Dordrecht: Springer)
	\bibitem[Aerts et al.(2003)]{2003Sci...300.1926A} Aerts, C., Thoul, A., Daszy{\'n}ska, J., et al.\ 2003, Science, 300, 1926. doi:10.1126/science.1084993
	\bibitem[Aerts et al.(2006)]{2006ApJ...642..470A} Aerts, C., Marchenko, S.~V., Matthews, J.~M., et al.\ 2006, \apj, 642, 470. doi:10.1086/500800
	\bibitem[Balona \& Ozuyar(2020)]{2020MNRAS.493.5871B} Balona, L.~A. \& Ozuyar, D.\ 2020, \mnras, 493, 5871. doi:10.1093/mnras/staa670
	\bibitem[Baran \& Koen(2021)]{2021AcA....71..113B} Baran, A.~S. \& Koen, C.\ 2021, \actaa, 71, 113. doi:10.32023/0001-5237/71.2.3
	\bibitem[Briquet et al.(2012)]{2012MNRAS.427..483B} Briquet, M., Neiner, C., Aerts, C., et al.\ 2012, \mnras, 427, 483. doi:10.1111/j.1365-2966.2012.21933.x
	\bibitem[Burssens et al.(2019)]{2019MNRAS.489.1304B} Burssens, S., Bowman, D.~M., Aerts, C., et al.\ 2019, \mnras, 489, 1304. doi:10.1093/mnras/stz2165
	\bibitem[Burssens et al.(2023)]{2023NatAs...7..913B} Burssens, S., Bowman, D.~M., Michielsen, M., et al.\ 2023, Nature Astronomy, 7, 913. doi:10.1038/s41550-023-01978-y
	\bibitem[Dziembowski \& Pamiatnykh(1993)]{1993MNRAS.262..204D} Dziembowski, W.~A. \& Pamiatnykh, A.~A.\ 1993, \mnras, 262, 204. doi:10.1093/mnras/262.1.204
	\bibitem[Dziembowski \& Pamyatnykh(2008)]{2008MNRAS.385.2061D} Dziembowski, W.~A. \& Pamyatnykh, A.~A.\ 2008, \mnras, 385, 2061. doi:10.1111/j.1365-2966.2008.12964.x
	\bibitem[Eze \& Handler(2024)]{2024ApJS..272...25E} Eze, C.~I. \& Handler, G.\ 2024, \apjs, 272, 2, 25. doi:10.3847/1538-4365/ad39c5
	\bibitem[Gaia Collaboration et al.(2018)]{2018A&A...616A...1G} Gaia Collaboration, Brown, A.~G.~A., Vallenari, A., et al.\ 2018, \aap, 616, A1
	\bibitem[Gaia Collaboration et al.(2021)]{2021A&A...649A...1G} Gaia Collaboration, Brown, A.~G.~A., Vallenari, A., et al.\ 2021, \aap, 649, A1. doi:10.1051/0004-6361/202039657
	\bibitem[Gaia Collaboration et al.(2016)]{2016A&A...595A...1G} Gaia Collaboration, Prusti, T., de Bruijne, J.~H.~J., et al.\ 2016, \aap, 595, A1. doi:10.1051/0004-6361/201629272
	\bibitem[Han et al.(2020)]{2020RAA....20..161H} Han, Z.-W., Ge, H.-W., Chen, X.-F., et al.\ 2020, Research in Astronomy and Astrophysics, 20, 161. doi:10.1088/1674-4527/20/10/161
	\bibitem[Handler et al.(2006)]{2006MNRAS.365..327H} Handler, G., Jerzykiewicz, M., Rodr{\'\i}guez, E., et al.\ 2006, \mnras, 365, 327. doi:10.1111/j.1365-2966.2005.09728.x
	\bibitem[Higgins \& Bell(2023)]{2023AJ....165..141H} Higgins, M.~E. \& Bell, K.~J.\ 2023, \aj, 165, 4, 141. doi:10.3847/1538-3881/acb20c
	\bibitem[Jenkins et al.(2010)]{2010ApJ...713L..87J} Jenkins, J.~M., Caldwell, D.~A., Chandrasekaran, H., et al.\ 2010, \apjl, 713, L87. doi:10.1088/2041-8205/713/2/L87
	\bibitem[Kurtz(2006)]{2006CoAst.147....6K} Kurtz, D.~W.\ 2006, Communications in Asteroseismology, 147, 6. doi:10.1553/cia147s6
	\bibitem[Labadie-Bartz et al.(2020)]{2020AJ....160...32L} Labadie-Bartz, J., Handler, G., Pepper, J., et al.\ 2020, \aj, 160, 32. doi:10.3847/1538-3881/ab952c
	\bibitem[Langer et al.(2020)]{2020A&A...638A..39L} Langer, N., Sch{\"u}rmann, C., Stoll, K., et al.\ 2020, \aap, 638, A39. doi:10.1051/0004-6361/201937375
	\bibitem[Lenz \& Breger(2005)]{2005CoAst.146...53L} Lenz, P. \& Breger, M.\ 2005, Communications in Asteroseismology, 146, 53. doi:10.1553/cia146s53
	\bibitem[Mazumdar et al.(2006)]{2006A&A...459..589M} Mazumdar, A., Briquet, M., Desmet, M., et al.\ 2006, \aap, 459, 589. doi:10.1051/0004-6361:20064980
	\bibitem[Miglio et al.(2007)]{2007CoAst.151...48M} Miglio, A., Montalb{\'a}n, J., \& Dupret, M.-A.\ 2007, Communications in Asteroseismology, 151, 48. doi:10.1553/cia151s48
	\bibitem[Montgomery \& Odonoghue(1999)]{1999DSSN...13...28M} Montgomery, M.~H. \& Odonoghue, D.\ 1999, Delta Scuti Star Newsletter, vol. 13, p.28, 13
	\bibitem[Moskalik \& Dziembowski(1992)]{1992A&A...256L...5M} Moskalik, P. \& Dziembowski, W.~A.\ 1992, \aap, 256, L5
	\bibitem[Murphy et al.(2019)]{2019MNRAS.485.2380M} Murphy, S.~J., Hey, D., Van Reeth, T., et al.\ 2019, \mnras, 485, 2, 2380. doi:10.1093/mnras/stz590
	\bibitem[Paxton et al.(2011)]{2011ApJS..192....3P} Paxton, B., Bildsten, L., Dotter, A., et al.\ 2011, \apjs, 192, 3. doi:10.1088/0067-0049/192/1/3
	\bibitem[Paxton et al.(2013)]{2013ApJS..208....4P} Paxton, B., Cantiello, M., Arras, P., et al.\ 2013, \apjs, 208, 4. doi:10.1088/0067-0049/208/1/4
	\bibitem[Paxton et al.(2015)]{2015ApJS..220...15P} Paxton, B., Marchant, P., Schwab, J., et al.\ 2015, \apjs, 220, 15. doi:10.1088/0067-0049/220/1/15
	\bibitem[Paxton et al.(2018)]{2018ApJS..234...34P} Paxton, B., Schwab, J., Bauer, E.~B., et al.\ 2018, \apjs, 234, 34. doi:10.3847/1538-4365/aaa5a8
	\bibitem[Paxton et al.(2019)]{2019ApJS..243...10P} Paxton, B., Smolec, R., Schwab, J., et al.\ 2019, \apjs, 243, 10. doi:10.3847/1538-4365/ab2241
	\bibitem[Pedersen et al.(2021)]{2021NatAs...5..715P} Pedersen, M.~G., Aerts, C., P{\'a}pics, P.~I., et al.\ 2021, Nature Astronomy, 5, 715. doi:10.1038/s41550-021-01351-x
	\bibitem[Pedersen et al.(2020)]{2020MNRAS.495.2738P} Pedersen, M.~G., Escorza, A., P{\'a}pics, P.~I., et al.\ 2020, \mnras, 495, 2738. doi:10.1093/mnras/staa1292
	\bibitem[Pigulski \& Pojma{\'n}ski(2008)]{2008A&A...477..917P} Pigulski, A. \& Pojma{\'n}ski, G.\ 2008, \aap, 477, 917. doi:10.1051/0004-6361:20078581
	\bibitem[Ramspeck et al.(2001)]{2001A&A...378..907R} Ramspeck, M., Heber, U., \& Moehler, S.\ 2001, \aap, 378, 907. doi:10.1051/0004-6361:20011246
	\bibitem[Ricker et al.(2015)]{2015JATIS...1a4003R} Ricker, G.~R., Winn, J.~N., Vanderspek, R., et al.\ 2015, Journal of Astronomical Telescopes, Instruments, and Systems, 1, 014003. doi:10.1117/1.JATIS.1.1.014003
	\bibitem[Sadowski et al.(2008)]{2008ApJ...676.1162S} Sadowski, A., Belczynski, K., Bulik, T., et al.\ 2008, \apj, 676, 1162. doi:10.1086/528932
	\bibitem[Shi et al.(2021a)]{2021AJ....161...46S} Shi, X.-D., Qian, S.-B., Li, L.-J., et al.\ 2021, \aj, 161, 46. doi:10.3847/1538-3881/abccd7
	\bibitem[Shi et al.(2021b)]{2021MNRAS.505.6166S} Shi, X.-D., Qian, S.-B., Li, L.-J., et al.\ 2021, \mnras, 505, 6166. doi:10.1093/mnras/stab1657
	\bibitem[Shi et al.(2021c)]{2021PASP..133e4201S} Shi, X.-D., Qian, S.-B., Li, L.-J., et al.\ 2021, \pasp, 133, 054201. doi:10.1088/1538-3873/abf32a
	\bibitem[Shi et al.(2022)]{2022ApJS..259..50S} Shi, X.-D., Qian, S.-B., \& Li, L.-J.\ 2022, \apjs, 259, 50. doi:10.3847/1538-4365/ac59b9
	\bibitem[Shi et al.(2023a)]{2023ApJS..265...33S} Shi, X.-D., Qian, S.-B., Zhu, L.-Y., et al.\ 2023, \apjs, 265, 33. doi:10.3847/1538-4365/acba91
	\bibitem[Shi et al.(2023b)]{2023ApJS..268...16S} Shi, X.-D., Qian, S.-B., Zhu, L.-Y., et al.\ 2023, \apjs, 268, 16. doi:10.3847/1538-4365/ace88c
	\bibitem[Shi et al.(2024)]{2024ApJS..271...28S} Shi, X.-D., Qian, S.-B., Zhu, L.-Y., et al.\ 2024, \apjs, 271, 1, 28. doi:10.3847/1538-4365/ad1f66
	\bibitem[Stankov \& Handler(2005)]{2005ApJS..158..193S} Stankov, A. \& Handler, G.\ 2005, \apjs, 158, 193. doi:10.1086/429408
	\bibitem[Stassun et al.(2018)]{2018AJ....156..102S} Stassun, K.~G., Oelkers, R.~J., Pepper, J., et al.\ 2018, \aj, 156, 102. doi:10.3847/1538-3881/aad050
	\bibitem[Vink et al.(2001)]{2001A&A...369..574V} Vink, J.~S., de Koter, A., \& Lamers, H.~J.~G.~L.~M.\ 2001, \aap, 369, 574. doi:10.1051/0004-6361:20010127
	\bibitem[Wenger et al.(2000)]{2000A&AS..143....9W} Wenger, M., Ochsenbein, F., Egret, D., et al.\ 2000, \aaps, 143, 9. doi:10.1051/aas:2000332
	\bibitem[Yoon et al.(2010)]{2010ApJ...725..940Y} Yoon, S.-C., Woosley, S.~E., \& Langer, N.\ 2010, \apj, 725, 940. doi:10.1088/0004-637X/725/1/940

	
\end{thebibliography}
\end{document}